\documentstyle[preprint,prb,aps]{revtex}
\begin{document}
\draft
\title{First order transition from ferromagnetism to antiferromagnetism in Ce(Fe$_{0.96}$Al$_{0.04}$)$_2$: a magnetotransport study}
\author{Kanwal Jeet Singh, Sujeet Chaudhary$^*$, M.K. Chattopadhyay, M.A. Manekar, S.B. Roy and P. Chaddah}
\address{Low Temperature Physics Laboratory, Centre for Advanced Technology, Indore 452 013, India}
\date{\today}
\maketitle
\begin{abstract}
The magnetotransport behaviour is investigated in detail across the first order magnetic  phase transition from ferromagnetic to antiferromagnetic state in polycrystalline Ce(Fe$_{0.96}$Al$_{0.04}$)$_2$ sample. The study clearly brings out various generic features associated with a first order transition, viz., hysteresis, phase coexistence, supercooling and superheating, presence and limits of the metastable regimes. These results of magnetotransport study exhibit and support all the interesting thermomagnetic history effects that were observed in our earlier dc-magnetisation study on the same sample. Most notable here is the initial (or virgin) resistivity vs. field curve lying outside the hysteretic "butterfly shaped" magnetoresistivity loops obtained on cyclying the magnetic field between high enough positive and negative strengths. These findings, bearing one-to-one similarity with the  data obtained in their magnetic counterpart (i.e., dc-magnetisation), have been ascribed an origin due to the arresting of this first order transition kinetics at low temperature and high magnetic field.
\end{abstract}
\pacs{}
\newpage

\section{Introduction}
The C15-Laves phase CeFe$_2$ compound, which crystalizes in the form of cubic MgCu$_2$-type structure, is an unstable ferromagnet with a Curie temperature, T$_C$ $\approx$235K. On doping with small concentrations of metals like Co, Ru, Ir, Al, Os, Re etc., at the Fe-site, the ferromagnetic (FM) behaviour disappears at low temperature, and the pseudobinary system Ce(Fe$_{1-x}$R$_x$)$_2$ exhibits a stable antiferromagnetic (AFM) phase \cite{1,2,3,4,5,6,7,8,9,10,11} below a certain temperature which we shall henceforth refer to as T$_N$. There exists an extensive amount of research work on this interesting magnetic system. Various experimental tools have been employed to know the true magnetic state of Ce sublattice and the instablities related with the Fe-sublattice. Studies based on neutron scattering \cite{12,13,14}, magnetic circular X-ray dichroism \cite{15,16,17}, magneto-optic Kerr effect \cite{18}, high energy spectroscopy (including x-ray photoemission, absorption, ultraviolet) \cite{19,20}, magnetic Compton scattering \cite{21}, specific heat \cite{22}, magnetisation \cite{8,9,10,11,23,24}, magnetotransport \cite{25,26,27,28} ac-susceptibility \cite{29,30,31,32} etc. have been reported in pure and doped CeFe$_2$ alloys. While these studies were mainly associated with the understanding of the magnetic instability in Ce(Fe$_{1-x}$R$_x$)$_2$, we have recently started probing the exact nature of the magnetic phase transition (AFM to FM), whcih can be driven by high enough magnetic field (H) even at temperatures (T) much below T$_N$ $\approx$ 100K. Our preliminary attempt on this front started initially with the ac-susceptibility study \cite{33} and investigation of the DC-magnetisation behaviour \cite{34} in the Ru- and Ir-doped CeFe$_2$ alloys, namely, the polycrystalline Ce(Fe$_{0.95}$Ir$_{0.05}$)$_2$ and Ce(Fe$_{0.93}$Ru$_{0.07}$)$_2$ samples. In these studies (based primarily on arguments of hysteresis and phase coexistence) the AFM to FM transition is found to be of first order in nature. Due to the very narrow hysteretic regime in these Ru and Ir doped CeFe$_2$ alloys,  we decided to investigate the Ce(Fe$_{0.96}$Al$_{0.04}$)$_2$ alloy wherein the AFM to FM transition is relatively gradual \cite{6,7}; this sample also provided us a distinctly wider H-T phase space (permitted by the maximum magnetic field (i.e., 5.5T) in our SQUID magnetometer) for a more comprehensive study of the nature of magnetic transition in this system. This revealed further interesting DC-magnetisation behaviour \cite{35,36} which along with some of our preliminary results of magnetotransport measurements \cite{36}, have led us to clearly establish the first order nature of the AFM to FM transition in the Ce(Fe$_{0.96}$Al$_{0.04}$)$_2$ alloy. The present paper deals with our detailed magnetotransport study on the same sample used in Refs.[35] and [36].
Although preliminary results of magnetotransport study on Ce(Fe$_{0.96}$Al$_{0.04}$)$_2$ have been published in (Ref.[36]), in this paper we (i) extend our investigation regime in the the resistivity $\rho$ vs. T hysteresis data to higher H; (ii) present $\rho$ vs. H hysteresis data to lower T; (iii) measure  minor hysteresis loops in $\rho$ vs. T; (iv) compare $\rho$ vs. T data obtained in zero-field cooled and field cooled cooling histories with that recorded while field cooled warming of the sample; and (v) study $\rho$ by following a more contrived history dependence of the external magnetic field. The present work shows that the Ce(Fe$_{0.96}$Al$_{0.04}$)$_2$ compound can be taken as a model system where one can observe most of the generic features related with a first order phase transition. We interpret all these results in the light of a first order transition and its associated behaviour (e.g., hysteresis, phase coexistance, supercooling and superheating, field dependent thermomagnetic history of magnetotransport behaviour). In addition, a comparison of the present work on Ce(Fe$_{0.96}$Al$_{0.04}$)$_2$ alloy has been made with the existing work on perovskite type manganese oxide compounds \cite{37,38,39} at appropriate places.

\section{Experimental}
The Ce(Fe$_{0.96}$Al$_{0.04}$)$_2$ alloy sample employed in the present magnetotransport study belongs to the same batch of samples used earlier in the study of bulk magnetic and transport properties \cite{6}, and neutron measurements\cite{12}. Details of the sample preparation and characterization can be found in Ref.[6]. A commercial cryostat (Oxford Instruments Inc., UK) with a maximum magnetic field of 16T is being used for carrying out the four probe resistivity measurement as a function of temperature and magnetic field H applied transverse to the measuring current. The isofield $\rho$ vs. T data was recorded with the following different protocols:\\
1. Zero Field Cooling (ZFC): The sample was first cooled from above T$_C$ down to lowest temperature of measurement in zero field, then subjected to the magnetic field, and $\rho$(T) data was recorded during warming of the sample. \\
2. Field Cooled Cooling (FCC): The sample was subjected to desired magnetic field strength above its T$_C$, and the $\rho$(T) data was recorded while cooling the sample.\\
3. Field Cooled Warming (FCW): The resistivity of the sample was measured as a function of T during warming of the sample which had been earlier field cooled to the lowest investigated temperature.

\section{Results}
Figure 1 shows the T-dependence of resistivity for the Ce(Fe$_{0.96}$Al$_{0.04}$)$_2$ sample in absence of magnetic field recorded during the initial cooling cycle starting from 290K, as well as during the subsequent warming cycle. During the initial decrease in temperature, while the paramagnetic (PM) to FM transition is reflected in the change of slope in $\rho$ vs. T plot at about 200K, the rise of $\rho$ is observed below about 84K due to a second magnetic phase transition from FM to AFM ordered state. This second transition gets completed at about 35K (while cooling), below which the resistivity once again decreases with decrease in temperature. For comparison, we present in the inset of Fig.1 a dc-magnetisation, M vs. T plot for the same sample in 2mT magnetic field. The three magnetic phases marked as PM, FM and AFM in different T-regimes complements our present results of the resistivity measurement. During the thermal cycling (i.e., warming and cooling), the resistivity behaviour exhibited distinct hysteresis across the FM to AFM transition (see main panel of Fig.1).

In Figs.2a-d, $\rho$ vs. T plots for magnetic field strengths of 0.7T, 1.5T, 2.5T and 4.0T, are respectively shown for the three different histories of application of magnetic field, e.g., ZFC, FCC and FCW ($\rho$ vs. T plots in ZFC and FCC protocols for other fields; 0.5T, 2.0T and 3.0T can be found in Fig.4 of Ref.[36].). As is evident from Figs.2a-d, the resistivity plots for all the three protocols - ZFC, FCC and FCW are strongly affected by the strength of the applied magnetic field across the FM-AFM transition. Note that both the onset and completion of AFM-FM transitions are shifted to lower temperatures in the presence of magnetic field. This shift is again found to be strongly field dependent. Compared to a value of $\approx$93K for completion of AFM-FM transition while warming the sample in the absence of field (see Fig.1), the same transition gets completed only by $\approx$46K when a magnetic field of 4T is applied (see Fig.2d). The magnetic field has another interesting effect in the low temperature AFM-state. Above a certain value of H (i.e., 0.5T), the FCC resistivity stays below the ZFC resistivity at low temperatures (see also Fig.4 of Ref.[36]). At any temperature, this difference of resistivity is field-dependent, increasing with the applied magnetic field. Furthermore starting at 4.5K, the FCW-data overlapped with the FCC-data upto some field-dependent temperature. Above this temperature, the FCW-curve continues to rise (in contrast to the falling FCC-curve) until it merges with the ZFC-curve at some slightly higher T. Thus the $\rho$ vs. T curves under the FCC and FCW protocols show the downturn at different temperatures. It is to be noted here that above the merger point of FCW- and ZFC-$\rho$(T) curves, the $\rho$(T) curve of the FCC-protocol lies distinctly below that recorded in the FCW-protocol. We stress here that FCW-protocol is basically a minor hysteresis loop of second kind (as would be described in next para) initiated from the lowest investigated temperature. 

Figure 3 shows the results obtained across what are called the "minor hysteresis loops" (MHL's) in absence of external magnetic field. Two kinds of MHL's are recorded. In the first kind, the sample is initially warmed from sufficiently low temperature, and then from a predetermined temperature the sample is cooled back. The MHL initiated at 44.5K from the warming cycle (Fig.3) is an example of this kind. In the second kind, the sample is initially cooled (from well above T$_C$) upto a predetermined temperature, and then it is warmed. The MHL's inititated on the cooling curve at 60K, 75K and 84K in the Fig.3 are examples of this kind. It can be clearly seen that the two MHLs initiated near the onset and completion of AFM-FM transition, respectively at 44.5K and 84K are reversible. However, the MHL's initiated from cooling curve at 60K and 75K followed an altogether different path. They finally merged with the warming curve (initiated from 4.5K, and would be henceforth referred to as envelope warming curve). 

The isothermal $\rho$ vs. H data was also recorded at various temperatures ranging from 2K to 120K. We show in Figs.4a and 4b, the $\rho$ vs. H data for the lowest temperature (i.e., 2K) and at 100K, respectively. Within our accuracy of measurements, the $\rho$ vs. H plots for T $\geq$ 100K were found to be linear right from 0T. As pointed out previously \cite{36}, the $\rho$(H) behaviour at lower temperatures is quite interesting. A small rise in $\rho$ with initial increase of H (within the AFM regime) along the virgin-curve is observed only at 2K (see Fig.4a).

Finally, we present in Fig.5, the results of resistivity measurements in 0.7T magnetic field during a warming cycle starting from 4.5K temperature point which has been approached using an altogether different protocol. In this protocol, the final state (i.e., 4.5K,0.7T) is being prepared following two steps. The first step involved the FCC protocol in 2.5T from above T$_C$ to 4.5K. Subsequent to this, in a second step, the field was reduced isothermally from 2.5T to 0.7T (these two steps are indicated as path-I in the inset of Fig.5). For the purpose of comparison, the $\rho$(T) plots for FCC and FCW protocols at 0.7T (corressponding to cooling and warming along the path-II as indicated in the inset of Fig.5) are also shown in Fig.5. With the initial reduction of field from 2.5T to 0.7T field at 4.5K, the resistivity value showed a jump (as indicated by a vertical arrow from point marked "A" at 4.5K in Fig.5) to a value (marked as point "B", Fig.5) which was smaller than the $\rho$(4.5K,0.7T) value recorded in any of the conventional ZFC, FCC or FCW measurement protocols for H=0.7T. However, on subsequent warming of the sample, the resistivity was found to increase as shown in Fig.5 until it finally merged with the $\rho$ vs. T curve in 0.7T corressponding to the FCW protocol.   

\section{Discussion}
\subsection{First order AFM to FM transition and the associated hysteresis}
Based on the $\rho$ vs. T data, we first argue that with increase in temperature and/or field, the low temperature low field AFM phase transforms to FM phase through a FOT. The first indication of a FOT comes in the form of hysteresis in resistivity which is observed in both the temperature (Figs.1 and 2) and field cyclings (Fig.4). During the cooling of the sample, the onset of the rise of resistivity starts with the nucleation of AFM phase at $\approx$84K (Fig.1). This upturn in resistivity basically occurs due to the appearence of magnetic superzones at the onset of FM-AFM transition \cite{6,7} (i.e., at T$_N$). Further cooling converts more and more fraction of the sample from FM state to AFM state, with the result that the entire sample transforms into AFM state below about 35K. During the warming cycle, the decrease in resistivity at the transition is slightly delayed compared to the cooling cycle, and starts at around 48K. This decreasing $\rho$(T) across the transition is again associated with the more and more fraction of sample getting converted into FM-state. Within the hysteretic resistivity regime, the difference in resistivity values at any temperature is associated with the relative fraction of  coexisting phases. (i.e., at any temperature within the hysteretic $\rho$(T) regime, this fraction has a different value for the cooling and warming cycles.) 

It is to be emphasized here that the transition from AFM to FM state during the warming cycle completes at distinctly higher temperature ($\approx$ 93K) than the onset of FM to AFM transition ($\approx$ 84K) during a cooling cycle. Similar to this, the onset of AFM to FM transition during warming cycle takes place at a relatively higher temperature (i.e., $\approx$ 48K) compared to completion of FM to AFM transition at $\approx$ 35K during the cooling cycle. The difference in the two temperatures at both the ends of the hysteretic regime is an indication that the FM to AFM transition in the present investigations of Ce(Fe$_{0.96}$Al$_{0.04}$)$_2$ is first order in nature. As we shall see later, the effect of the applied field is to affect these various transition temperatures and the magnitude of thermal hysteresis in the resistivity (Fig.2). 

We wish to state here that the FM to AFM transition in the present case of Ce(Fe$_{0.96}$Al$_{0.04}$)$_2$ sample is quite broad on temperature axis in comparison to relatively sharp FM to AFM transition observed in single crystal manganese oxide perovskite samples \cite{37,38,39}. The broadening of a first order transition with the disorder in the sample is predicted theoretically \cite{40} way back in 1979, and very recently the magneto-optic imaging study on the vortex matter of single crystal Bi-2212 sample has shown how the disorder leads to a distribution of transition temperatures and/or fields across the solid to liquid melting transition, which leads to heterogeneous nucleation across the transition \cite{41}. We believe that a similar distribution of transition temperatures and/or fields exists in the present case of polycrystalline sample of Ce(Fe$_{0.96}$Al$_{0.04}$)$_2$. Accordingly, the transition from one magnetic phase to another magnetic phase in the present work on Ce(Fe$_{0.96}$Al$_{0.04}$)$_2$ should be designated here by a band, instead of a transition line on H-T phase space.

\subsection{Minor Hysteresis Loops and Phase-coexistence}
The study of minor hysteresis loops \cite{42} (MHL's) has been recently recognised as an important experimental technique to study the phenomenon of phase coexistence across the FOT in more detail \cite{33,35,43,44}. Our magnetotransport results (Fig.3) provide support in favor of phase coexistence across the foregoing AFM to FM transition. Consider the MHL's that were initiated well inside the hysteretic regime. They showed a strong irreversible behaviour. For example, the MHL initiated on cooling cycle at 60K, during its course (i.e., while warming) did not come reversibly along the $\rho$(T) of cooling curve due to the already transformed AFM phase at 60K. Instead, a slight increase in resistivity with T is observed along this initial course of MHL due to the temperature dependence of transformed AFM phase akin to $\rho$(T) behaviour in the low temperature AFM regime below 35K. On further warming along the MHL, this increasing resistivity behaviour changes back to decreasing resistivity behaviour due to more and more AFM to FM conversion with the result that the MHL finally merges with the envelope warming cycle at about 70K. The evidence of (a) phase coexistence and (b) the hysteresis due to the different fractions of FM and AFM phases is clearly obvious from the fact that the resistivity of the sample measured at any temperature within the irreversible portion of MHL, is drastically different than that recorded along either the envelope cooling or warming cycle at the same temperature.

On the other hand, the reversible behaviour of MHL initiated from the warming-cycle at 44.5K (a temperature at which a distinct thermal hysteresis in resistivity exists between the envelope cooling and warming cycles) comes due to the fact that the onset of nucleation of FM has not taken place until 44.5K during the warming cycle. Thus while the sample is purely in the AFM state until 44.5K during warming cycle, a finite fraction of the sample continued to remain ferromagnetic upto 35K while cooling. With the same argument, the MHL initiated from the cooling cycle at 84K which took a overlapping course along the envelope cooling cycle indicated the absence of any AFM phase at or above 84K during the initial cooling cycle. This state of the sample (i.e. fully FM) at 84K during the cooling cycle (as well as along the MHL under consideration) is in sharp contrast to that along the envelope warming cycle where the sample contains a finite fraction of AFM phase (which is yet to transform into FM state) for even for T$>$84K as indicated by $\rho$ vs. T plot for 84K$<$T$<$93K (Fig.3). We note here that a clear co-existence of FM and AFM phases was also observed in zero field neutron measurement \cite{12} in a substantial T-regime below the onset of the phase transition in Ce(Fe$_{1-x}$Al$_x$)$_2$ samples with 0.2$\leq$x$\leq$0.08. 

\subsection{Metamagnetic transition and associated metastabilities}
The $\rho$ vs. H plots for T $\geq$ 100K exhibited typical ferromagnetic response i.e., negative magnetoresistance which increases in magnitude with field right from H = 0. However, at lower temperatures, the $\rho$ vs. H plots in the virgin as well as field increasing envelope cycle showed field induced AFM to FM transition, commonly referred to as "metamagnetic transition", resulting from spin-flipping in the AFM phase at high enough fields \cite{45}. The decrease of resistivity due to onset of ferromagnetism at H$_m$(T) continues (due to conversion of more and more antiferro-phase into ferro-phase) until some higher field after which the resistivity decreases linearly with the field indicating the usual negative magnetoresistance ($\propto$ H) behaviour of the fully transformed high field FM state of the entire sample. We note here that the positive magnetoresistance behaviour (i.e., $\rho$ increasing with increase in H) of the AFM state is distinctly observed only in the low temperature $\rho$ vs. H plot recorded at 2K (Fig.4a). This positive sign of magnetoresistance in a AFM state is consistent with the previous reports \cite{46} on Fe-Mn-Cr ternary alloys. 

Along the H-decreasing envelope branch of a $\rho$ vs. H curve (for T $\leq$ 80K), the entire sample remains in the form of high field FM phase down to relatively lower field values compared to those along either the virgin curve or the H-increasing envelope curve. This is expected across a first order transition \cite{47} as a sign of metastable behaviour due to (a) the supercooling of the high field FM phase below the transition field, and (b) superheating of low field AFM phase above the transition field, when field is cycled across the transition field (i.e., H$_m$ line or more prescise H$_m$-band). Further, one expects that for some value of H below H$_m$(T) the resistivity behaviour of the sample should once again become AFM like. However, the results of Fig.4a (this work) and Fig.2 of Ref.[36]) clearly show that this is not the case with Ce(Fe$_{0.96}$Al$_{0.04}$)$_2$. This is so because resistivity along the H-decreasing envelope curve is anomalous since on cycling the field back to H=0, the resistivity value corresponding to the low field stable AFM phase [i.e., that of H-increasing envelope curve (in case of T$>$10K) and/or virgin curve (in case of T$\leq$10K)] is not completely restored back, with the result that a "open hysteresis loop" is being obtained. Similar behaviour was found in virgin curve initiated in negative H direction for T$\leq$10K (Ref.[36]). This indicates that FM phase persists in some finite quantity even at zero field after being cycled from +H$_{max}$ or -H$_{max}$. It is to be recalled here that although similar open hysteresis loops were also reported for Nd$_{0.5}$Sr$_{0.5}$MnO$_3$ (Ref.[36]), the hysteresis loops in case of Pr$_{0.5}$Sr$_{0.5}$MnO$_3$ did not display any such open hysteresis loop (Ref.[37]). 

At lower temperatures, the anomalous open hysteresis loop in the $\rho$(H) behaviour is manifested in the form of another anomaly below 10K, in that the virgin resistivity curve lie outside the butterfly shaped (envelope) hysteresis loop (e.g., Fig.4a, the resistivity along the virgin $\rho$(H) curve remains (upto H$_m$ or so) distinctly above both the envelope H-increasing as well as H-decreasing $\rho$ vs. H curves.). In case of magnetisation measurement, this anomaly is observed in the form of virgin M(H) curve lying outside the butterfly (envelope) hysteresis loop \cite{35,36}. We would come back to this anomalous result below H$_m$ a little later. We are unaware of any such comparison based on the resistivity measurements in all the three protocols (i.e., along the virgin, field-increasing envelope, and field-decreasing envelope curve) exists for either Nd$_{0.5}$Sr$_{0.5}$MnO$_3$, Nd$_{0.25}$Sm$_{0.25}$Sr$_{0.5}$MnO$_3$ and/or Pr$_{0.5}$Sr$_{0.5}$MnO$_3$ (Refs.[37-39]). 

\subsection{Limits of metastability: effect of magnetic field on the temperature dependence of resistivity}
It is important here to recall that there exists a limit of metastability \cite{47,48}, below (above) the first order transition line (in our case, the H$_m$(T) or T$_N$(H)) upto which one can supercool (superheat) the high (low) field phase. Outside these two bounds (designated as T$^*$(H) for supercooling, and T$^{**}$(H) for superheating) on either side of the transition line, no metastable behaviour, and hence no superheating and/or supercooling can be observed. (Recall that like T$_N$(H), as stated earlier these two bounds T$^*$(H) and T$^{**}$(H) should be represented by the bands in the present case of polycrytalline Ce(Fe$_{0.96}$Al$_{0.04}$)$_2$ sample.). It is also known that the metastable region widens with increase (decrease) of field (temperature). Natural question at this moment is - do we see such features of FOT in the present $\rho$(H,T) data of Ce(Fe$_{0.96}$Al$_{0.04}$)$_2$ sample? 

The comparison of the resistivity behaviour recorded on the Ce(Fe$_{0.96}$Al$_{0.04}$)$_2$ sample under different protocols (ZFC, FCC, and FCW) has turned out to be very interesting, and displays almost all the above mentioned generic features of a first order phase transition. We find that the magnetic field has a very drastic effect on the $\rho$(T) behaviour, e.g., with increasing magnetic field, \\
(a) the completion of FM to AFM transition while cooling the sample (which marks the lower limiting value of T$^*$(H)-band, and is inferred by the temperature below which the FCC and FCW-resistivities merge) is suppressed much faster than the decrease in the onset of AFM to FM transition temperature while warming the sample (i.e., the upper limiting value of T$^{**}$(H)).\\
(b) the hysteretic regime is substantially enhanced in temperature across the transition, and \\
(c) most importantly, the low temperature reversible $\rho$(T) regime (with respect to the  overlapping ZFC and FCC) disappears by H$\geq$0.7T.

These results ((a) and (b) above) clearly imply that with the increase in H, the lower limit of metastability, T$^*$(H) band  (below which one should see the reversible response of the stable AFM phase) gets suppressed (towards lower temperature) even faster in comparison to T$^{**}$(H) band, and the metastable regime (existing between lower limiting value of T$^*$(H) band and the upper limiting value of T$^{**}$(H) bands) thus widens with the decrease  in T or with increase in H. This is further supported from the isothermal $\rho$ vs. H data, where the hysteretic field regime drastically increased at low temperatures (Fig.4a). Based on these T dependence of  resistivity in differents fields, a phase diagram using the mid point of T$^*$(H)- and T$^{**}$(H)-band has been shown in our preliminary work \cite{36}.

Along the FCW curve, the resistivity rising past this merger point of FCW and FCC exhibits the superheating of low temperature AFM phase. The FCW curve finally merges with the ZFC curve at a temperature where the supercooled (metastable) FM-phase formed during the FCC-protocol vanishes. Above this field dependent merger temperature of FCW and ZFC resistivities, the sample comprises of three fractions; \\
1. stable transformed/nucleated FM phase, \\
2. superheated metastable AFM phase, and \\
3. stable AFM phase which has not yet transformed in FM-phase becuase of the distributuion in the T$_N$'s in the sample.

It is now quite evident that the present study of $\rho$(T) under different histories of applied magnetic field (i.e., ZFC, FCC, and FCW) in Ce(Fe$_{0.96}$Al$_{0.04}$)$_2$ has an edge over the previous work on Nd$_{0.5}$Sr$_{0.5}$MnO$_3$,  Nd$_{0.25}$Sm$_{0.25}$Sr$_{0.5}$MnO$_3$ and/or Pr$_{0.5}$Sr$_{0.5}$MnO$_3$, where resistivity data is compared only during FCC with FCW \cite{37,38,39}.  

From the thermomagnetic irreversibility increasing with the field as observed in the DC-magnetisation data of the same Ce(Fe$_{0.96}$Al$_{0.04}$)$_2$ sample \cite{35,36}, it was argued that the hysteresis in the magnetisation vs. field was entirely due to the first order nature of the magnetic transition rather than arising due to pinning of the domians formed in the FM state. All the results of present magnetotransport study are in excellent agreement with those observed in the DC-magnetisation studies \cite{35}. This is so because, the hysteresis observed in the present bulk magnetotransport behaviour can not be again related with the domain formation, since the domain size is normally much greater than the mean free path of the carriers \cite{46}.   

\subsection{Anomalous aspects of thermomagnetic history effects across the AFM to FM transition}

We once again recall here that in case of first order transition, the free energy barrier between the stable AFM phase and the metastable FM phase ceases to exists just below the T$^*$(H), and any infinitesimal fluctuations can drive the entire sample to stable AFM below the T$^*$(H) line/band \cite{47}. But, a few questions still remain unanswered in this picture: \\
1. Why does not the resistivity of the sample recorded in FCC and FCW protocol restore back to that recorded in ZFC protocol at lower temperatures (e.g., at 5K) for H $\geq$ 0.7T? (We tend to believe that similar feature would appear in case of resistivity behaviour in Nd$_{0.5}$Sr$_{0.5}$MnO$_3$ and Pr$_{0.5}$Sr$_{0.5}$MnO$_3$.) In our magnetisation studies on same Ce(Fe$_{0.96}$Al$_{0.04}$)$_2$ sample \cite{35,36}, this anomaly is reflected in the observation that the FC magnetisation curve M$^{FC}$(T) lie above the M$^{ZFC}$(T). \\
2. Why does the virgin $\rho$ vs. H curve lie outside the envelope butterfly $\rho$(H) loop at lower temperatures, T$\leq$10K? \\ 
3. Why is the full AFM resistivity not distinctly restored at H=0 on the envelope H-decreasing $\rho$ vs. H curve?

It is once again to be remembered that structural transition in Ce(Fe$_{1-x}$Al$_x$)$_2$ system for 0.2$\leq$x$\leq$0.08 from cubic to rhombohedral accompanies the AFM/FM transition \cite{12}. We now argue here that every structural transition is characterized by a characteristic relaxation time, which increases with decreasing temperature due to reduction in the displacive motion of the atoms \cite{49}. A typical example is the supercooled liquid-to-glass transition. It is our conjecture that in the present case of Ce(Fe$_{0.96}$Al$_{0.04}$)$_2$ sample, this characteristic relaxation time at low temperature may become much larger than the actual experimental time scale with the result that kinetics of transition gets hindered. Within this picture, the anomalies enumerated above can be explained as follows. When the temperature is lowered, some high temperature high field metastable FM phase remains frozen-in resulting in the arrest of FM to AFM transition at low temperature. In the $\rho$ vs. H cycling, due to the low temperature, a finite amount of supercooled FM phase continues to exist even when the applied field is reduced to zero. (Note that the magnetic field also induces the structural transition in Ce(Fe$_{0.96}$Al$_{0.04}$)$_2$ (Ref.[4]).) This metastable FM phase is then carried over when the direction of field is changed, and results in an H-decreasing (increasing) envelope curve that lie below the virgin curve recorded in the negative (positive) H direction. As expected, such hysteresis loops are not seen at higher temperatures ($>$10K).
 
Another evidence supporting the presence of hindered kinetics which, if not considered properly, might tempt one to draw false conclusion about the lower limit of metastable behaviour, comes from the results of Fig.5. As discussed above, the initial field cooling of the sample in a magnetic field of 2.5T results into a finite fraction of high temperature FM phase existing at 4.5K due to the hindered kinetics as mentioned in the preceeding para. Subsequent to this FC protocol, the isothermal (i.e., at 4.5K) field reduction from 2.5T to 0.7T yielded a resistivity value which is significantly smaller than either of $\rho$(4.5K,0.7T,ZFC) and/or $\rho$(4.5K,0.7,FCC/FCW), indicating as if the metastable behaviour in the sample existed at or below 4.5K in 0.7T field. This we believe is due to the hindered kinetics because while cooling the sample in 2.5T a large fraction (compared to 0.7T, FCC state at 4.5K) of supercooled high temperature FM phase is frozen, i.e., its transformation to stable AFM phase at T$^*$(H) gets arrested. As a result, the resistivity of the sample at 4.5K after the field is reduced to 0.7T is significantly smaller than the FCC/FCW resistivity in 0.7T at 4.5K. On warming the sample, this anomalously low resistivity of the sample at 4.5K increases fast (as the sample gains energy in the warming process), and the resistivity merges at about 20K with that of kinetically hindered resistivity curve of 0.7T. We thus say that compared to field cooling in 0.7T to 4.5K (i.e., path II in inset of Fig.5), the kinetics of the FM to AFM transition experienced relatively more hindrance when the sample was field cooled to the same 4.5K, 0.7T point but following path 'I'.

It is to be noted here that results obtained follwoing path-I experimental protocol (Fig.5) are similar to the results of magnetic annealing effect reported in the Cr-doped manganite crystals of Nd$_{0.5}$Ca$_{0.5}$Mn$_{1-y}$Cr$_y$O$_3$ (See Fig.2b of ref.[50]). [We believe that the value of magnetisation at 5K in their sample at the annealing field of 7T (which is not shown in Ref.[50]) would be much larger than the value obtained after field was reduced to 0.5T at 5K (See Fig.2b of Ref.[50]).] The authors of ref.[50] explained their data on the basis of random field quenching due to Cr-substitution which produced the FM-microembryos. Our experimental results of magnetotransport studies (Fig.5) are ascribed with arresting of the first order transition process at low temperature and high magnetic field. We further wish to point here that we have confirmed the magnetotransport results of Fig.5 by DC-magnetisation measurement (which are not shown here for the sake of concizeness) following similar path-I as indicated in the inset of Fig.5.      

In the Ref.[36], the hindrance to the kinetics of the transition has been indicated on the H-T phase diagram through a (H$_k$,T$_k$) band (see Fig.3b of Ref.[36]). The reader is referred to Fig.3c of Ref.[36] and the related text for a more illustrative understanding of hindered kinetics in the present case of Ce(Fe$_{0.96}$Al$_{0.04}$)$_2$.

\section{Conclusion}
The magnetotransport behaviour across the first order phase transition from low field and low temperature antiferromagnetic state to high field and high temperature ferromagnetic state has been studied in Ce(Fe$_{0.96}$Al$_{0.04}$)$_2$ polycrystalline sample. This study on Ce(Fe$_{0.96}$Al$_{0.04}$)$_2$ has clearly demonstrated the various generic features of a first order phase transitions, viz., hysteresis, phase coexistence, field dependent limits and existence of the metastable supercooled and superheated phases across the transition boundary in a magnetic system. These results of magnetotransport study not only supported our previous work of dc-magnetisation study on the same sample, but also provided new results extending to relatively higher magnetic field and low temperature regime on the H-T phase space. We found some unusal magnetoresistance behaviour (most interesting one is the virgin resistivity curve lying outside the envelope resistivity loop) at lower temperatures. Based on the observed magnetotransport behaviour using an unconventional history of application of magnetic field, we have ascribed the origin of these unusal features with the arresting of the transition at lower temperature. 
\\
\\
{\bf Acknowledgements} \\ 
We thank R.K. Meena and Anil Chouhan for the assistance provided in the magnetotransport measurements.
\\
\\
$^*$ Corresponding author, E-mail\hspace{.25in}:\hspace{.5in}sujeetc@cat.ernet.in

\begin{figure}
\caption{Resistivity vs. temperature plot for Ce(Fe$_{0.96}$Al$_{0.04}$)$_2$ sample recorded during the initial cooling (filled square symbols) from above 290K to 4.5K, and subsequent warming cycle (open square symbols). The inset shows a temperature dependence of DC-magnetisation on the same sample recorded in a field of 2.0mT.}
\end{figure} 
\begin{figure}
\caption{Effect of magnetic field on the $\rho$ vs. T plots for Ce(Fe$_{0.96}$Al$_{0.04}$)$_2$ sample recorded under different measurement protocols, viz. ZFC (open square symbols), FCC (open triangle symbols) and FCW (filled triangle symbols); (a) 0.7T, (b)1.5T, (c) 2.5T, and (d) 4.0T.}
\end{figure}
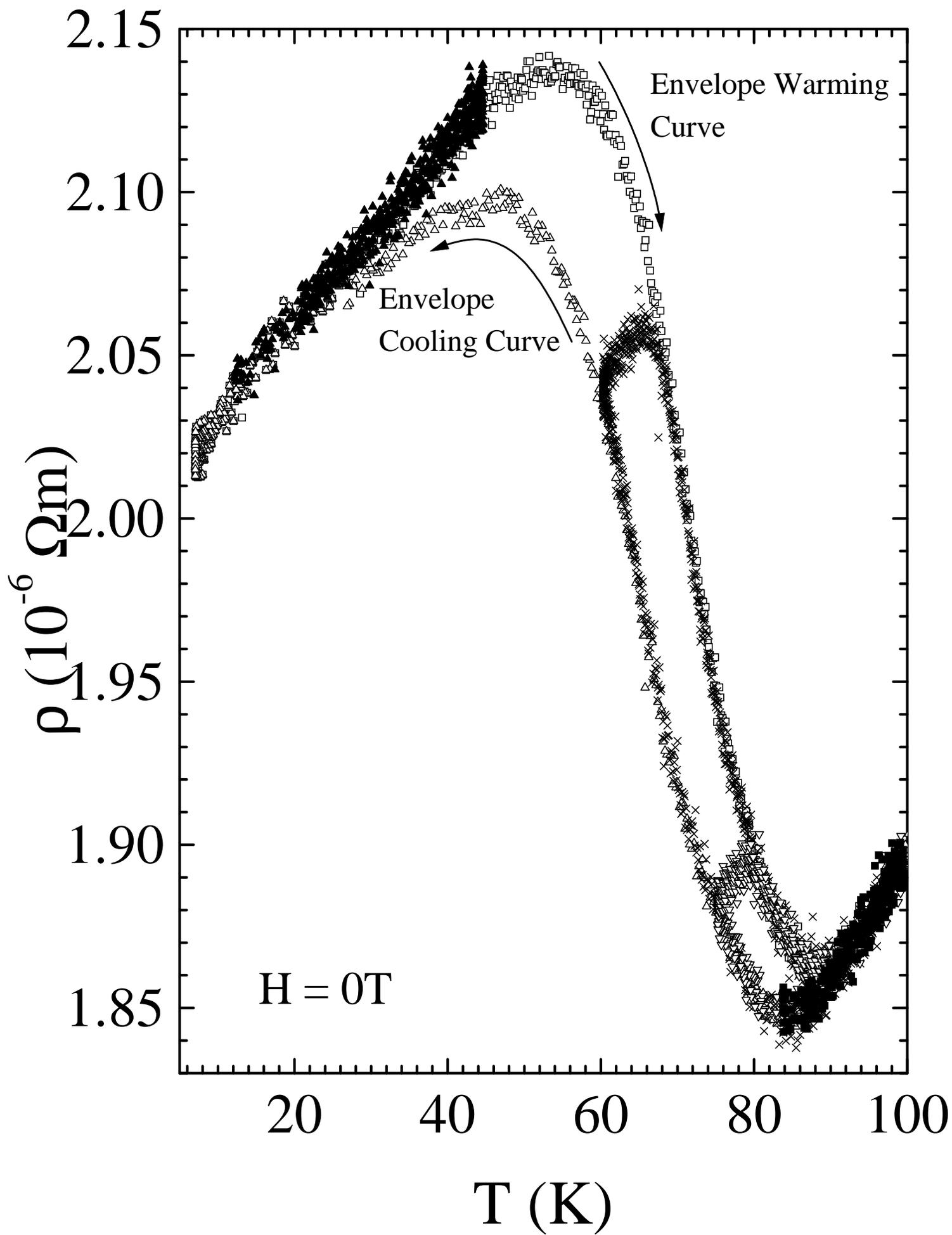
\begin{figure}
\caption{The minor hysteresis loops for Ce(Fe$_{0.96}$Al$_{0.04}$)$_2$ sample initiated from the envelope warming curve (open square symbols) at 44.5K (filled up-triangle symbols), and from the envelope cooling curve (open up-triangle symbols) at 60K (cross symbols), 75K (open down-triangle symbols) and 84K (filled square symbols). Refer text for more details.}
\end{figure}
\begin{figure}
\caption{The isothermal $\rho$ vs. H plots for Ce(Fe$_{0.96}$Al$_{0.04}$)$_2$ sample recorded at (a) 2K and (b) 100K. The three different histories of H and T in the $\rho$ vs. T behaviour (at 2K), i.e., virgin resistivity curve (starting from H=0 after initially cooling the sample to the desired temperature in zero field), envelope H-decreasing resistivity curve and the envelope H-increasing resistivity curve are shown respectively using the filled triangle, open square and open triangle symbols. See text for details.}
\end{figure}
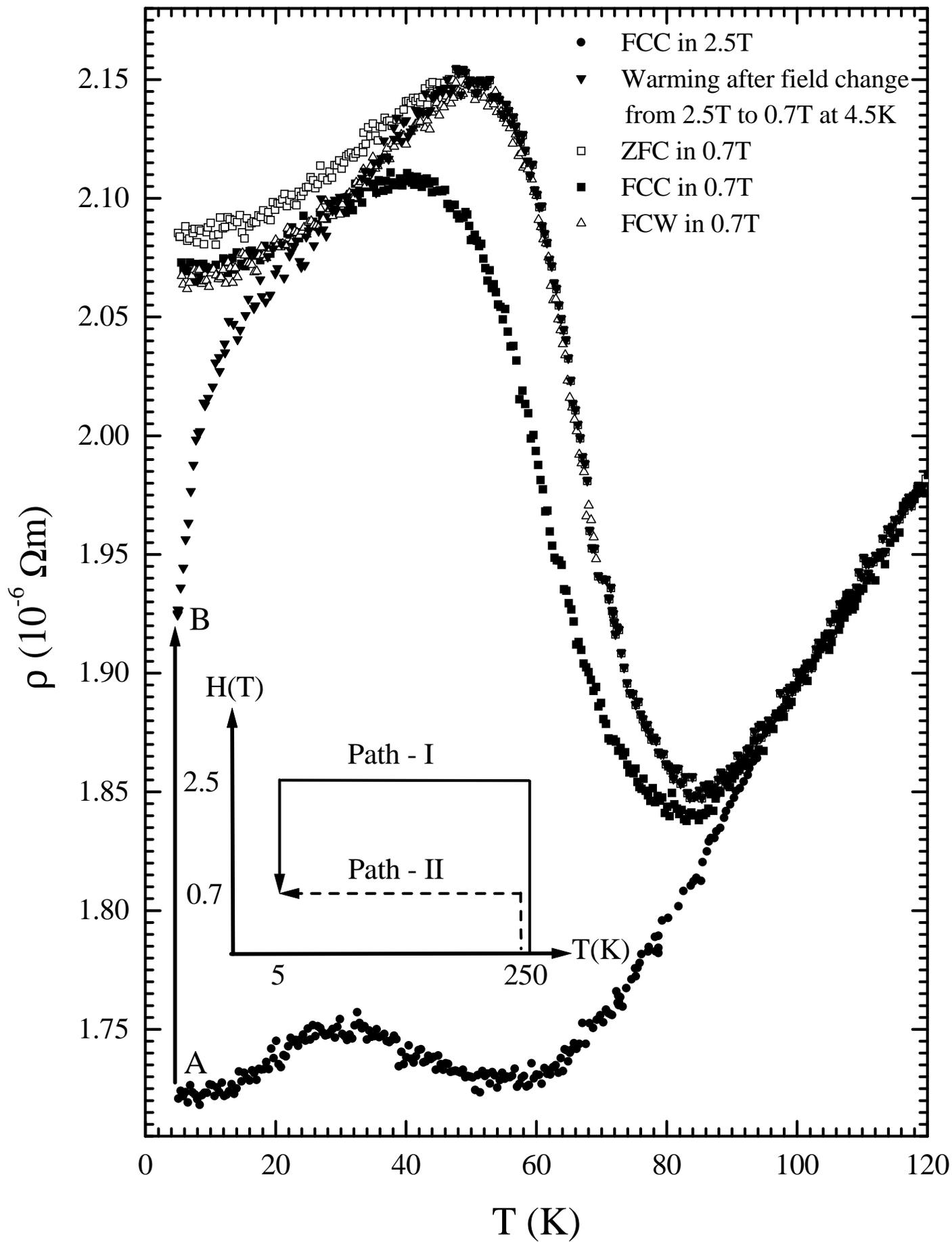
\begin{figure}
\caption{The $\rho$ vs. T plot (filled down-triangle) for Ce(Fe$_{0.96}$Al$_{0.04}$)$_2$ sample recorded during a heating cycle initiated from a starting point (0.7T,4.5K) represented as point 'B' in the main panel following path-I (shown in the inset). Path-I involved two steps. While the first step (basically a FCC protocol in 2.5T) is shown by filled circle symbols, the second step involving the isothermal field reduction from 2.5T to 0.7T is shown by an arrow (from point 'A' to point 'B' in the main panel of the figure). For the purpose of comparison, the $\rho$ vs. T plots recorded for ZFC (open square symbols), FCC (filled square symbols) and FCW (open up-triangle symbols) protocols for H=0.7T are also shown in the figure.}
\end{figure}

\begin{references}
\bibitem{1}D.F. Franceschini and S.F. Da Cunha, J. Magn. Magn. Mater. {\bf 52} (1985) 280.
\bibitem{2}A.K. Rastogi and A.P. Murani, {\it Theoretical and Experimental Aspects of Valence Fluctuations and Heavy Fermions} edited by L.C. Gupta and S.K. Malik (New York) Plenum (1987) p437.
\bibitem{3}S.B. Roy and B.R. Coles, J. Phys. F: Met. Phys. {\bf 17} (1987) L215.
\bibitem{4}S.B. Roy, S.J. Kennedy, and B.R. Coles, J. Physique Coll. {\bf 49} C8 (1988) 271.
\bibitem{5}A.K. Rastogi, G. Hilsher, E. Gratz and N. Pillmayr, J. Physique Coll. {\bf 49} C8 (1988) 277.
\bibitem{6}S.B. Roy and B.R. Coles, J. Phys.: Condens. Matter {\bf 1} (1989) 419.
\bibitem{7}S.B. Roy and B.R. Coles, Phys. Rev. B {\bf 39} (1989) 9360.
\bibitem{8}S. Radha, S.B. Roy, A.K. Nigam and Girish Chandra, Phys. Rev. B {\bf 50} (1994) 6866.
\bibitem{9}A.K. Rajarajan, S.B. Roy and P. Chaddah, Phys. Rev. B {\bf 56} (1997) 7808.
\bibitem{10}C.C. Tang, Y.X. Li, J. Du, G.H. Wu and W.S. Zhan, J. Phys.: Condens. Matter {\bf 11} (1999) 2027.
\bibitem{11}H. Fukuda, H. Fujii and H. Kamura, Phys. Rev. B {\bf 63} (2001) 54405.
\bibitem{12}S.J. Kennedy and B.R. Coles, J. Phys.: Condens. Matter {\bf 2} (1990) 1213.
\bibitem{13}S.J. Kennedy, P.J. Brown and B.R. Coles, J. Phys.: Condens. Matter {\bf 5} (1993) 5169.
\bibitem{14} L. Paolasini, P. Dervenagas, P. Vulliet, J.-P. Sanchez, G.H. Lander, A. Hiess, A. Panchula and P. Canfield, Phys. Rev. B {\bf 58} (1998) 12117.
\bibitem{15}C. Giorgetti, S. Pizzini, E. Dartyge, A. Fontaine, F. Baudelet, C. Brouder, Ph. Bauer, G. Krill, S. Miraglia, D. Fruchart and J.P. Kappler, Phys. Rev. B {\bf 48} (1993) 12732.
\bibitem{16}J. Ph. Schille, F. Bertran, M. Finazzi, Ch. Brouder, J.P. Kappler and G. Krill, Phys. Rev. B {\bf 50} (1994) 2985.
\bibitem{17}A. Delobbe, A.-M. Dias, M. Finazzi, L. Stichaure, J.-P. Kappler and G. Krill, Europhys. Lett. {\bf 43} (1998) 320.
\bibitem{18}R.J. Lange, I.R. Fisher, P.C. Canfield, V.P. Antropov, S.J. Lee, B.N. Harmon and D.W. Lynch, Phys. Rev. B {\bf 62} (2000) 7084.
\bibitem{19}J. Chaboy, C. Piquer, L.M. Garcia, F. Bartlome, H. Wada, H. Maruyama and N. Kawamura, Phys. Rev. B, {\bf 62} (2000) 468.
\bibitem{20}T. Konishi, K. Morikawa, K. Kobayashi, T. Mizokawa, A. Fujimori, K. Mamiya, F. Iga, H. Kawanaka, Y. Nishihara, A. Delin and O. Ericksson, Cond-mat/000921 13 Sep 2000.
\bibitem{21}M.J. Cooper, P.K. Lawson, M.A.G. Dixon, E. Zukowski, D.N. Timms, F. Itoh, H. Sakurai, H. Kawata, Y. Tanaka and M. Ito, Phys. Rev. B {\bf 54} (1996) 4068.
\bibitem{22}H. Wada, T. Harada and M. Shiga, J. Phys.: Condens. Matter {\bf 9} (1997) 9347.
\bibitem{23}H.P. Kunkel, X.Z. Zhou, P.A. Stampe, J.A. Cowen and G.Williams,  Phys. Rev. B {\bf 53} (1996) 15099.
\bibitem{24}J. Deportes, D. Givord and K.R.A. Ziebeck, J. Appl. Phys. {\bf 52} (1981) 2074.
\bibitem{25}S. Radha, S.B. Roy and A.K. Nigam, J. Appl. Phys. {\bf 87} (2000) 1.
\bibitem{26}C.S. Garde, J. Ray and G. Chandra, Phys. Rev. B {\bf 42} (1990) 8643.
\bibitem{27}E. Gratz, E. Bauer, H. Nowotny, A.T. Burkov and M.V. Vedernikov, Solid State Commun. {\bf 69} (1989) 1007.
\bibitem{28}N. Ali and X. Zhang, J. Phys.: Condens. Matter {\bf 4} (1992) L351.
\bibitem{29}J. Eyon and N. Ali, J. Appl. Phys. {\bf 69} (1991) 5063.
\bibitem{30}X. Zhang and N. Ali, J. Appl. Phys. {\bf 75} (1994) 7128.
\bibitem{31}D. Wang, H.P. Kunkel and G. Williams, Phys. Rev. B {\bf 51} (1995) 2872.
\bibitem{32}R. Caciuffo, J.-C. Griveau, D. Kolberg, F. Wastin, D. Rinaldi, A. Panchula and P. Canfield, Phys. Rev. B {\bf 85} (1999) 6229.
\bibitem{33}M. Manekar, S.B. Roy and P. Chaddah, J. Phys.: Condens. Matter {\bf 12} (2000) L409.
\bibitem{34}M. Manekar, S. Chaudhary, M.K. Chattopadhyay, K.J. Singh, S.B. Roy and P. Chaddah, J. Phys.: Condens. Matter {\bf 12} (2000) 9645.
\bibitem{35}M. Manekar, S. Chaudhary, M.K. Chattopadhyay, K.J. Singh, S.B. Roy and P. Chaddah, communicated to J. Phys.: Condens. Matter (2001).
\bibitem{36}M. Manekar, S. Chaudhary, M.K. Chattopadhyay, K.J. Singh, S.B. Roy and P. Chaddah, Phys. Rev. B {\bf 64} (2001) 104416.
\bibitem{37}H. Kuwahara, Y. Tomioka, A. Asamitsu, Y. Moritomo and Y. Tokura, Science {\bf 270} (1995) 961.
\bibitem{38}Y. Tomioka, A. Asamitsu, Y. Moritomo, H. Kuwahara and Y. Tokura, Phys. Rev. Lett. {\bf 74} (1995) 5108.
\bibitem{39}Y. Tokura, H. Kuwahara, Y. Moritomo, Y. Tomioka, and A. Asamitsu, Phys. Rev. Lett. {\bf 76} (1996) 3184.
\bibitem{40}Y. Imry and M. Wrtis, Phys. Rev. B {\bf 19} (1979) 3580. 
\bibitem{41}A. Soibel, E. Zeldov, M. Rappaport, Y. Myasoedov, T. Tamegai, S. Ooi, M. Konczykowski and V.B. Geshkenbein, Nature {\bf 406} (2000) 283.
\bibitem{42}P. Chaddah, S.B. Roy, S. Kumar and K. Bhagwat, Phys. Rev. B {\bf 46} (1992) 11737.
\bibitem{43}S.B. Roy and P. Chaddah, Physica C {\bf 279} (1997) 70.
\bibitem{44}X. Granados, J. Fontcuberta and X. Obradors, Phys. Rev. B {\bf 46} (1992) 15683.
\bibitem{45}B.D. Cullity, in {\it Introduction to Magnetic Materials}, Addison-Wesley Publishing Company (1972) p240.
\bibitem{46}T.K. Nath and A.K. Majumdar, Phys. Rev. B {\bf 57} (1998) 10655.
\bibitem{47}P.M. Chaikin and T.C. Lubensky in {\it Principles of Condensed Matter Physics}, Cambridge University Press (1995).
\bibitem{48}P. Chaddah and S.B. Roy, Phys. Rev. B {\bf 60} (1999) 11926. 
\bibitem{49}P.G. Debenedetti in {\it Metastable Liquids}, Princeton University Press (1996).
\bibitem{50}T. Kimura, Y. Tomioka, R. Kumai, Y. Okimoto and Y. Tokura, Phys. Rev. Lett. {\bf 83} (1999) 3940.

\end{references}
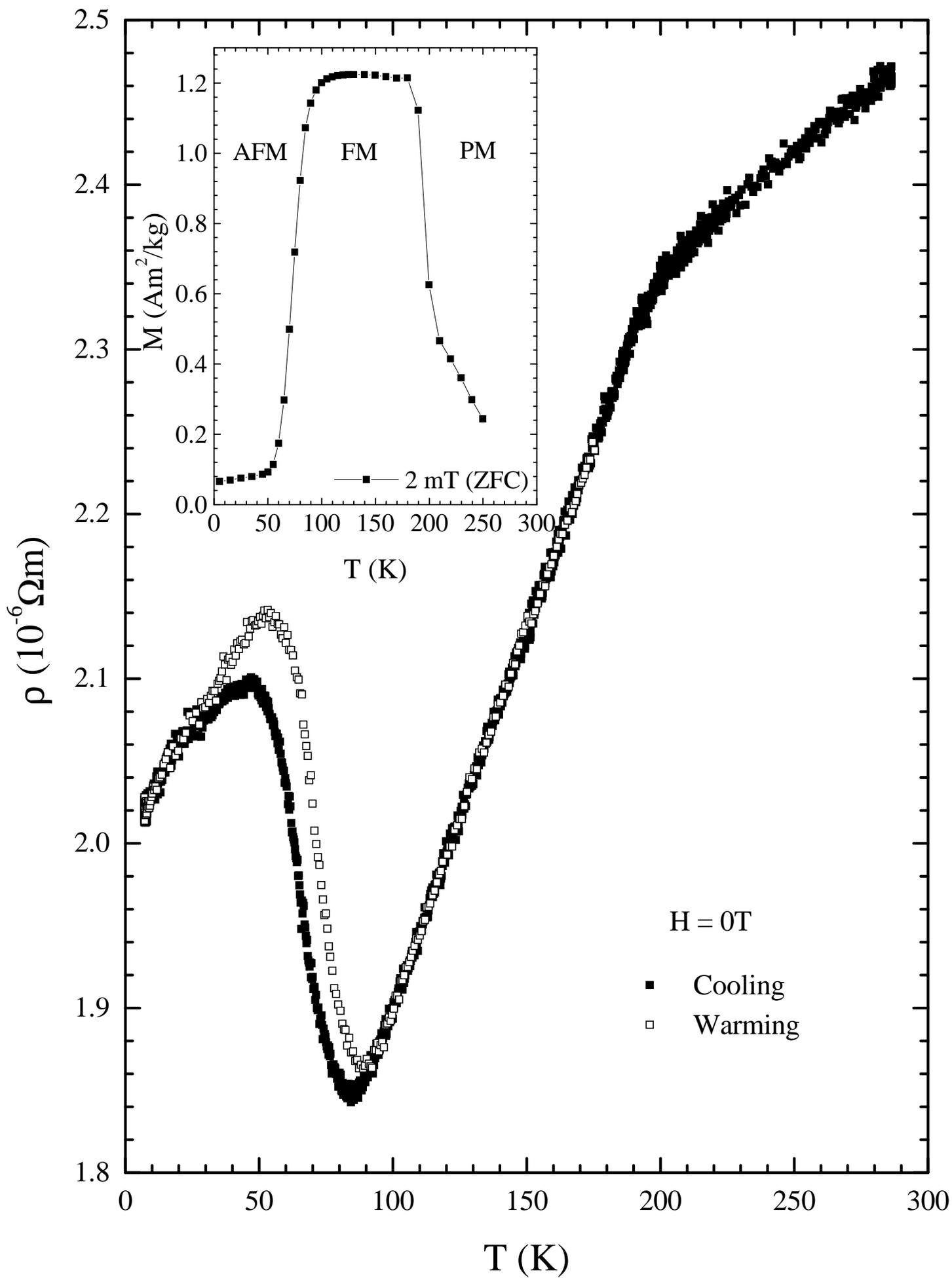
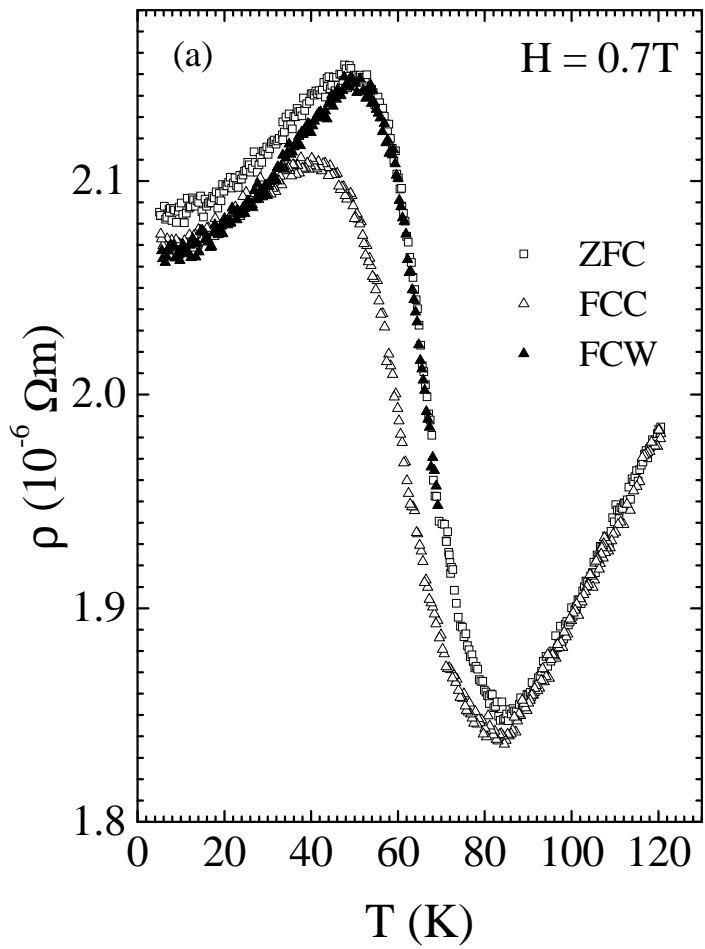
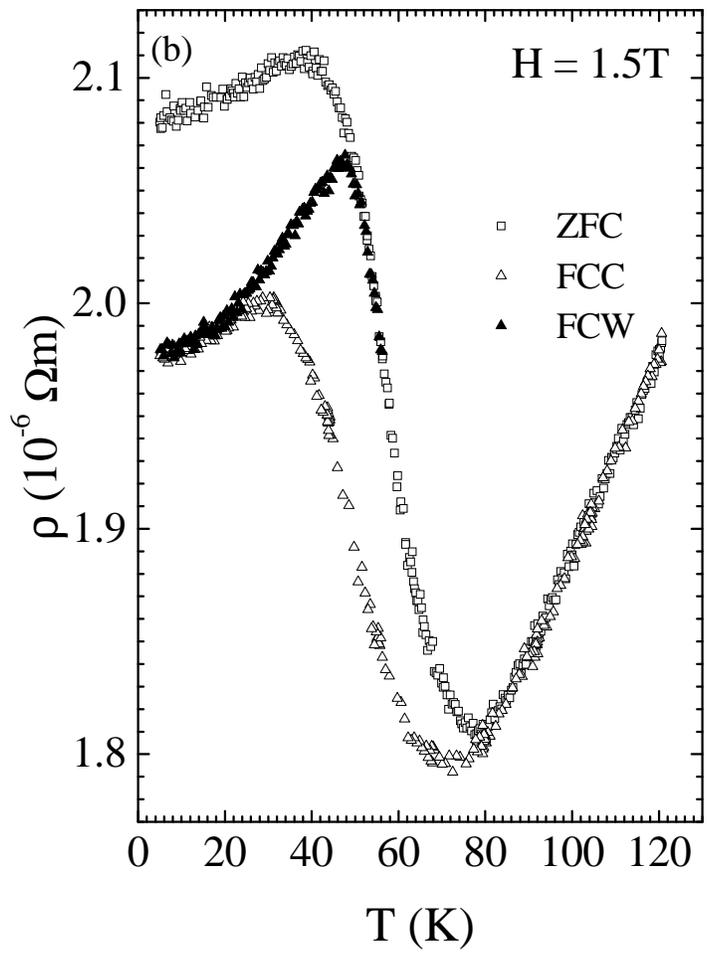
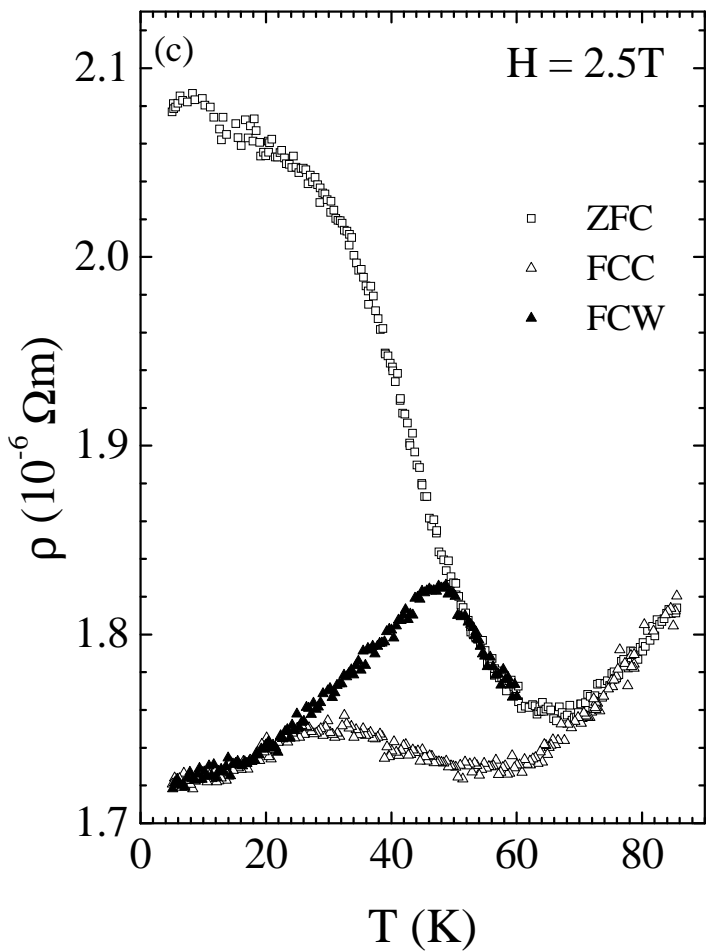
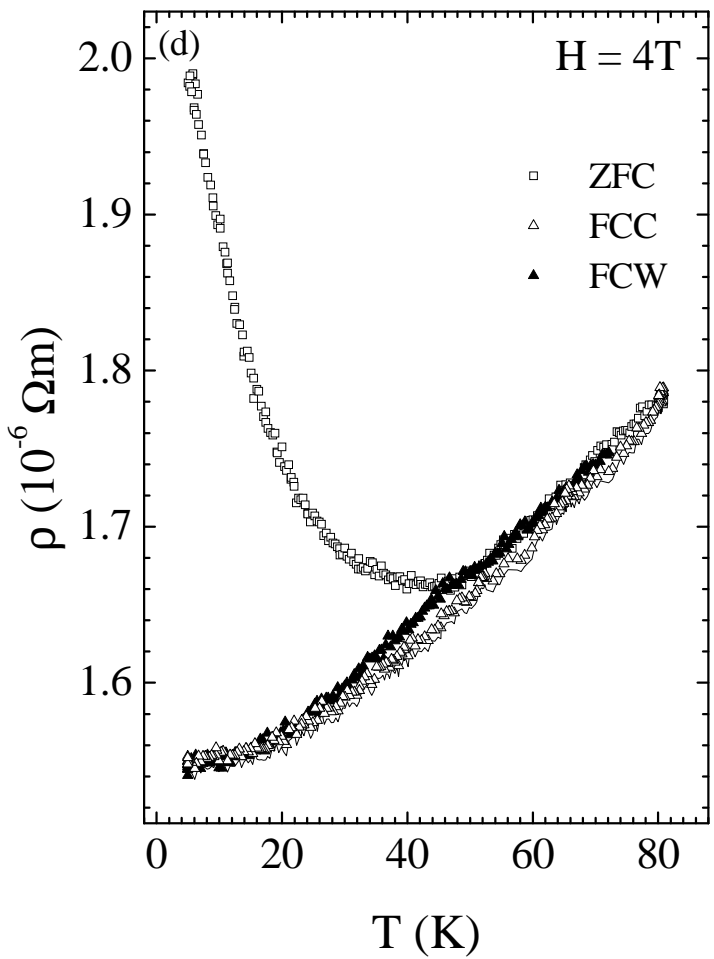
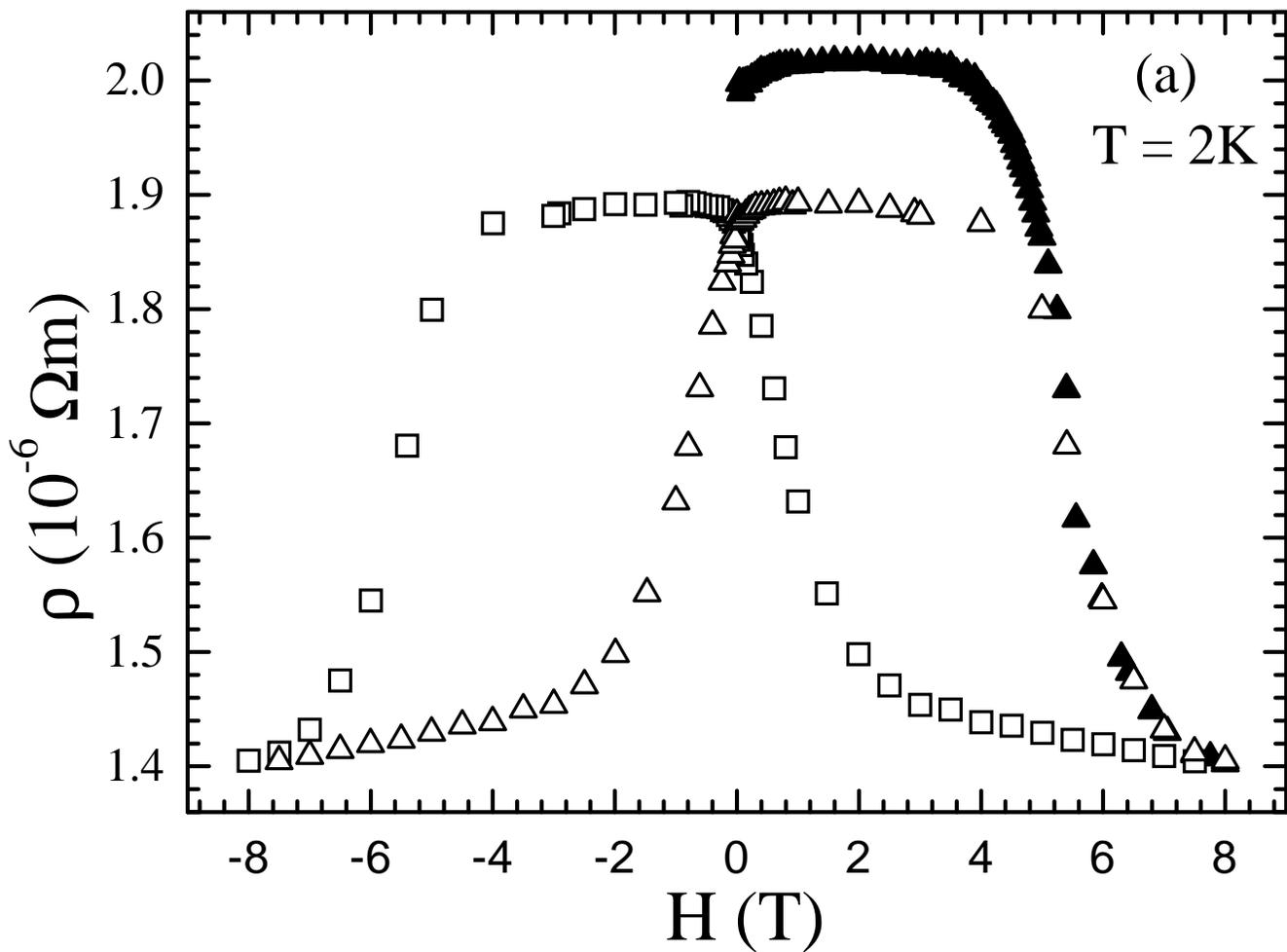
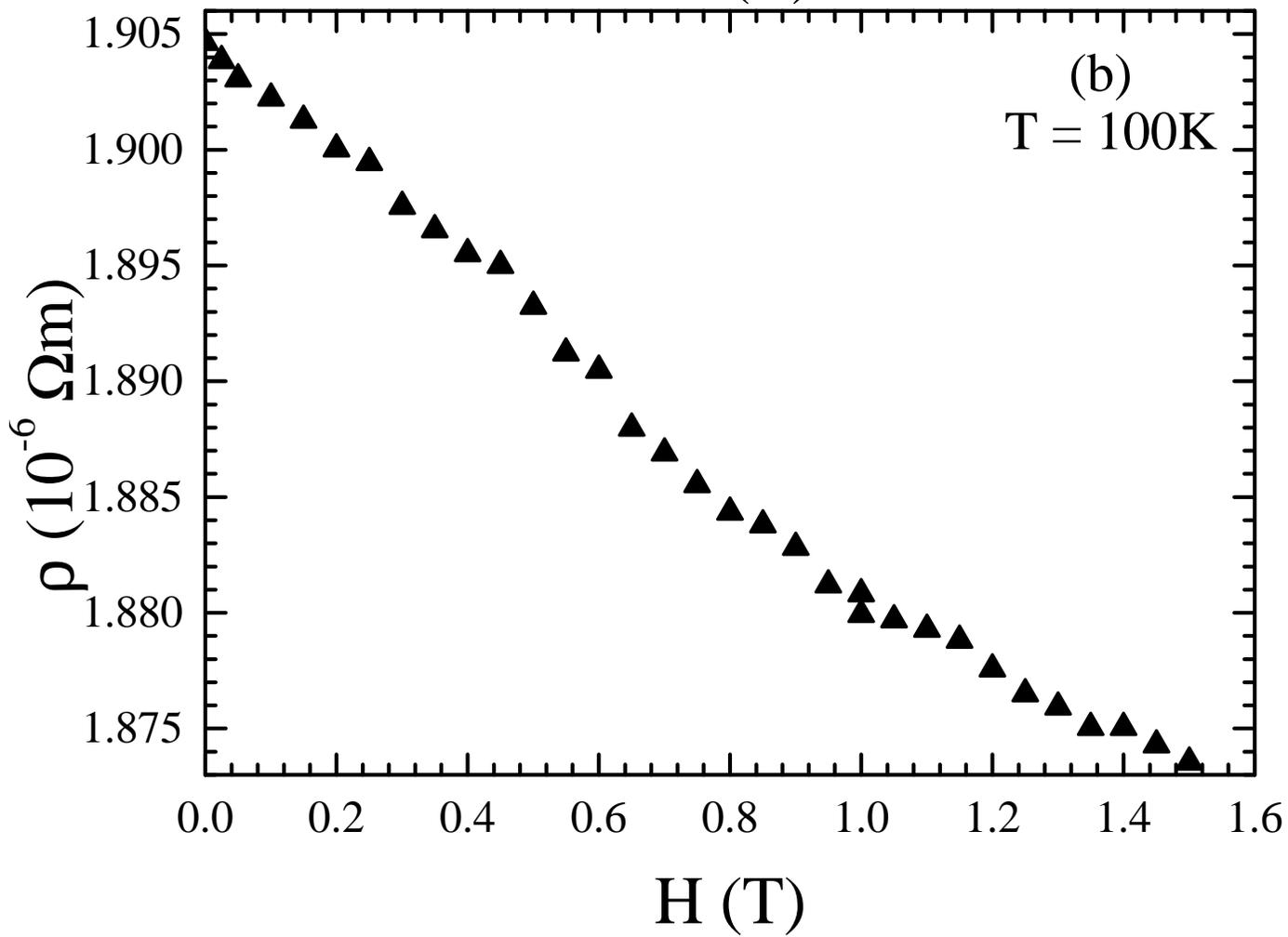
\end{document}